\documentclass{iaus}
\usepackage{graphicx}
\usepackage{natbib}

\title[Wind anisotropy and stellar evolution]{Wind anisotropy and stellar evolution}
\author[Cyril Georgy, Georges Meynet \& Andr\'e Maeder]{Cyril Georgy, Georges Meynet \and Andr\'e Maeder}
\affiliation{Geneva Observatory, University of Geneva, Maillettes 51 - CH 1290 Sauverny, Switzerland}
\pubyear{2008}
\volume{255}
\pagerange{}

\jname{Low-Metallicity Star Formation: From the First Stars to Dwarf Galaxies}
\editors{L.K. Hunt, S. Madden \& R. Schneider, eds.}

\begin{document}

\maketitle

\begin{abstract}
Mass loss is a determinant factor which strongly affects the evolution and the fate of massive stars. At low metallicity, stars are supposed to rotate faster than at the solar one. This favors the existence of stars near the critical velocity. In this rotation regime, the deformation of the stellar surface becomes important, and wind anisotropy develops. Polar winds are expected to be dominant for fast rotating hot stars.

These polar winds allow the star to lose large quantities of mass and still retain a high angular momentum, and  they modifie the evolution of the surface velocity and the final angular momentum kept in the star's core. We show here how these winds affect the final stages of massive stars, according to our knowledge about Gamma Ray Bursts. Computation of theoretical Gamma Ray Bursts rate indicates that our models have too fast rotating cores, and that we need to include an additional effect to spin them down. Magnetic fields in stars act in this direction, and we show how they modify the evolution of massive star up to the final stages.
\end{abstract}

\section{Effects of rotation on the stellar surface}
Rotation has a strong influence on the stellar surface. Indeed, it adds a centrifugal component to the gravity, which modifies the shape of the surface, and various quantities such as the effective temperature $T_\mathrm{eff}$ and the mass loss flux. These effects can be derived from the von Zeipel theorem \citep{vonZeipel1924a}, which is originally valid for conservative cases of angular momentum distribution, and was treated in the more general case of the so-called "shellular rotation"  by \citet{Maeder1999a}. This theorem gives the relation between the local flux $\mathbf{F}$ of the star and the local effective gravity:
\begin{equation}
\mathbf{F} = -\frac{L(P)}{4\pi G M_\star (P)}\mathbf{g_\mathrm{eff}}\left( 1 + \zeta(\theta)\right)
\end{equation}
\noindent where $L(P)$ is the luminosity on the isobar and $\mathbf{g_\mathrm{eff}}$ the local effective gravity. The two remaining terms are given by
\begin{eqnarray}
\zeta(\theta) &=& \left[ \left( 1-\frac{\chi_T}{\delta}\right)\Theta + \frac{H_T}{\delta}\frac{\mathrm{d}\Theta}{\mathrm{d}r}\right]P_2 (\cos(\theta))\\
M_\star &=& M\left( 1-\frac{\Omega^2}{2\pi G\rho_\mathrm{m}}\right)
\end{eqnarray}
\noindent Here, $\rho_\mathrm{m}$ is the internal average density, $M_\star$ represents the effective mass, modified by the rotation velocity $\Omega$, $\chi = 4acT^3 / (3\kappa\rho)$ is the thermal conductivity coefficient and $\chi_T$ is its partial derivative with respect to $T$, $\Theta = \frac{\tilde{\rho}}{\overline{\rho}}$ is the ratio of the horizontal density fluctuation to the average density on the isobar \citep{Zahn1992a}. $\delta$ is the thermodynamic coefficient $\delta = -\left( \partial\ln \rho / \partial\ln T\right)_{P,\,\mu}$ and $H_T$ is the temperature scale height. Generally the term $\zeta(\theta)$ is very small, and we can neglect it.

The total gravity at the surface of the star is given by $\mathbf{g_\mathrm{tot}} = \mathbf{g_\mathrm{eff}} + \mathbf{g_\mathrm{rad}} $ with $\mathbf{g_\mathrm{rad}} = \frac{\kappa\mathbf{F}}{c}$ is the term due to radiative forces and $\kappa$ the total Rossland mean opacity. The Eddington limit in a rotating star is defined by the vanishing of $\mathbf{g_\mathrm{tot}}$ and we find the limiting flux
\begin{equation}
\mathbf{F_\mathrm{lim}} = -\frac{c}{\kappa}\mathbf{g_\mathrm{eff}}\,.
\end{equation}

\noindent The Eddington factor at a given colatitude, which is given by the ratio of the local flux to the limiting flux, becomes \citep{Maeder1999a}
\begin{equation}
\Gamma_\Omega(\theta) = \frac{L(P)}{L_\mathrm{max}}\,\,\,\mathrm{with}\,\,\, L_\mathrm{max} = \frac{4\pi cGM}{\kappa(\theta)\left( 1 + \zeta(\theta)\right)}\left( 1-\frac{\Omega^2}{2\pi G\rho_\mathrm{m}}\right)\,\, .
\end{equation}
\noindent A first interesting consequence is that the maximum luminosity of a star, given by $\Gamma_\Omega(\theta) = 1$,  is lowered by rotation (compare $L_\mathrm{max}$ given above with the non-rotating one $L_\mathrm{max,\,no\,rot} = 
\frac{4\pi cGM}{\kappa}$).

Rotation has also a strong impact on the mass loss rate of the star. \citet{Maeder2000a} found the following relation between the mass loss rate at a given rotational velocity $\dot{M}(\Omega)$ and the non-rotating one:

\begin{equation}
\frac{\dot{M}(\Omega)}{\dot{M}(0)} =  \frac{\left( 1-\Gamma\right)^{\frac{1}{\alpha} - 1}}{\left[ 1-\frac{\Omega^2}{2\pi G\rho_\mathrm{m}}-\Gamma \right]^{\frac{1}{\alpha} - 1}}
\end{equation}
\noindent where $\Gamma = \frac{\kappa L}{4\pi cGM}$ is the "non-rotating" Eddington factor. We see that rotation will increase the total mass loss rate of the star.

Rotation has another effect on the mass loss rate: it is no long isotropic at the surface of a rotating star, but becomes colatitude-dependent. \citet{Maeder2000a} give the following local mass loss rate $\Delta\dot{M}$ per unit surface $\Delta\sigma$:
\begin{equation}
\frac{\Delta\dot{M}}{\Delta\sigma} \simeq \left( k\alpha\right)^\frac{1}{\alpha}\left( \frac{1-\alpha}{\alpha}\right)^\frac{1-\alpha}{\alpha}\left[\frac{L(P)}{4\pi GM_\star(P)}\right]^\frac{1}{\alpha}\frac{g_\mathrm{eff}}{\left(1-\Gamma_\Omega(\theta)\right)^{\frac{1}{\alpha}-1}}
\end{equation}
\noindent with $\kappa$ and $\alpha$ the force multiplier parameters. We neglect here the small effect of $\zeta(\theta)$. Rotation favors mass loss through the terms $M_\star$ and $\Gamma_\Omega$. Mass loss occurs preferentially where $g_\mathrm{eff}$ is small, \textit{i.e} at the poles. Then mass loss rate is thus varying as a function of the colatitude, producing the anisotropic wind phenomenon.

\begin{figure}
\begin{center}
\includegraphics[angle = 0, width=9 cm]{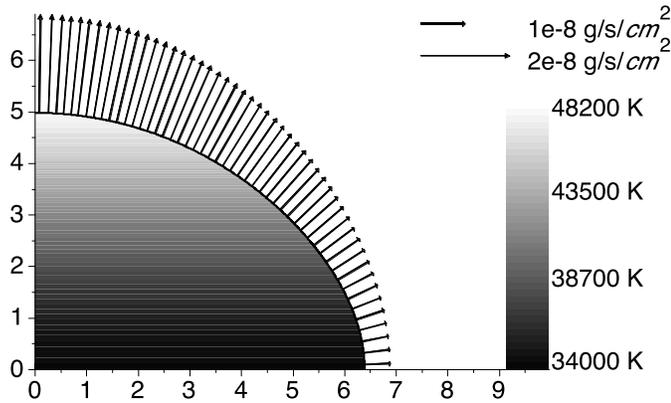}
\caption{Effects of rotation on a $20\,\mathrm{M}_\odot$ star at $Z=10^{-5}$ and with $\frac{\Omega}{\Omega_\mathrm{crit}}=0.95$. The star is seen equator-on, and the axis are in $\mathrm{R}_\odot$. The latitudinal variation of $T_\mathrm{eff}$ is shown (gray scale), and the arrows represents the masse flux at a given colatitude.}
\end{center}
\label{shape}
\end{figure}
Figure \ref{shape} shows various effects of the rotation at the surface of a $20\, \mathrm{M}_\odot$ star at a metallicity of $10^{-5}$ and at 95\% of the critical rotation velocity. We can first remark that the star becomes oblate, with an equatorial-to-polar radius ratio $\frac{R_\mathrm{eq}}{R_\mathrm{pol}}\simeq 1.3$. Then we note the variation of the effective temperature with respect to colatitude: $T_\mathrm{eff}$ at the pole is around $48000\,\mathrm{K}$, while it is only around $34000\,\mathrm{K}$ at the equator. Finally, we see that the mass loss flux is larger at the pole by a factor of $\sim 3.8$. This allows the star to lose  $10\%$ less angular momentum than if the same amount of mass was lost isotropically.

\section{Models without magnetic field and rate of GRB}

Let us briefly recall here the main assumptions of the so-called "collapsar" model for long-soft Gamma Ray Bursts \citep{Woosley1993a}. In order to produce such an event, the following conditions must be fulfilled:

\begin{itemize}
\item formation of a black hole;
\item enough angular momentum in the stellar core in order to form an accretion disk around the BH;
\item formation of a type Ic supernova (see \citet{Woosley2006a}).
\end{itemize}

\begin{figure}
\begin{center}
\includegraphics[angle=0,width=7 cm]{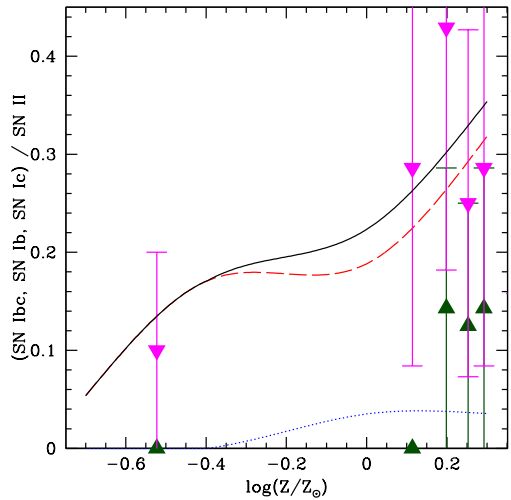}
\caption{SNIb/SNII (solid line), SNIb/SNII (dotted line) and SNIc/SNII (dashed line) ratios. The triangles are observed SNIb/SNII ratio at various metallicities and the upside down gray triangles observed SNIc/SNII ratio (extracted from \citet{Prieto2008a}).}
\label{SNRate}
\end{center}
\end{figure}

These three points agree with a fast rotating massive star. It is thus interesting to study the evolution of the rate of type Ic supernovae. Figure \ref{SNRate} shows the relative rates for type Ib and type Ic SNe, computed with rotating models without magnetic field. To distinguish between type Ib and type Ic, we use a criterion based on the amount of He ejected during the SN event: all models ejecting more than $0.55\,\mathrm{M}_\odot$ of helium and no hydrogen are considered as type Ib, the models ejecting less than $0.55\,\mathrm{M}_\odot$ of helium (and still no hydrogen) are considered to give birth to a type Ic SN event. The interesting curve for our purpose is the dashed one, representing the SN Ic / SN II ratio with respect to the metallicity. Wee see that the number fraction of type Ic supernova becomes higher at higher metallicity. This is well in agreement with the trend shown by the observed type Ic to type II SNe ratios given by \citet{Prieto2008a}. There are many observed evidences indicating that long-soft GRB occurs preferentially in metal-poor regions. For instance, \citet{Modjaz2008a}  find that GRB events appear at low metallicity: between $0.2 < Z/\mathrm{Z}_\odot < 0.7$. Thus only type Ic events in metal poor regions (or part of them) can occur simultaneously with a GRB event.

Moreover, if we consider our models at metallicities that are compatible with the observed GRB range of metallicity, we see that all the models producing a type Ic SN keep enough angular momentum in the core to fulfilled the collapsar model conditions (see \citet{Hirschi2005a}). We can thus determine the ratio of GRB event to the total number of core collapse SNe; we found that this rate is around $15\%$ for metallicities between $0.4$ and $0.7\,\mathrm{Z}_\odot$. In comparison, \citet{Podsiadlowski2004a} found an observational rate of $0.04\% - 8\%$, depending of the aperture angle of the bipolar jets produced during the GRB event. Our theoretical rate is therefore much larger than the observational one. These two facts lead to the conclusion that not all type Ic SNe produce a GRB. We have thus to find a way to reduce the number of GRB progenitor candidates, in order to reproduce the observational rate. One possibility is to introduce new ingredients in our models to extract more angular momentum from the core during the stellar life.

\section{Models with magnetic field and wind anisotropy}

Following \citet{Spruit2002a}, we have included in our models the effect of magnetic field amplified at the expense of the excess energy in the shear. As noted just above, this produces a strong coupling between the differentially rotating layers, and tend to build a solid--body rotation profile. Contrarily to the previous models, where the rotational velocity is only weakly coupled between the surface and the core, models with magnetic field have a strong coupling, and thus, the loss of angular momentum due to mass loss is quickly transmitted to the core. This implies a strong extraction of angular momentum during the evolution, particularly when the mass loss is strong at the surface (e.g. during the Wolf--Rayet phases).

To explore the combined effects of magnetic field and wind anisotropy, we computed two $60\,\mathrm{M}_\odot$ models at $Z = 0.002$ with  $\Omega / \Omega_\mathrm{crit} = 0.75$, with and without the treatment of the wind anisotropy. With respect to the work by \citet{Meynet2007a}, we have improved significantly the treatment of the wind anisotropy, allowing this treatment to apply even when the critical velocity is reached and checking very carefully that the sum of the angular momentum remaining in the star and the angular momentum lost in the wind remains constant all over the evolution (see Fig. \ref{Levol}, bottom panel). As we shall see below, this improvements lead to effects which although still important are less pronounced as in \citet{Meynet2007a}.

\begin{figure}
\begin{center}
\includegraphics[angle=0,width=9cm]{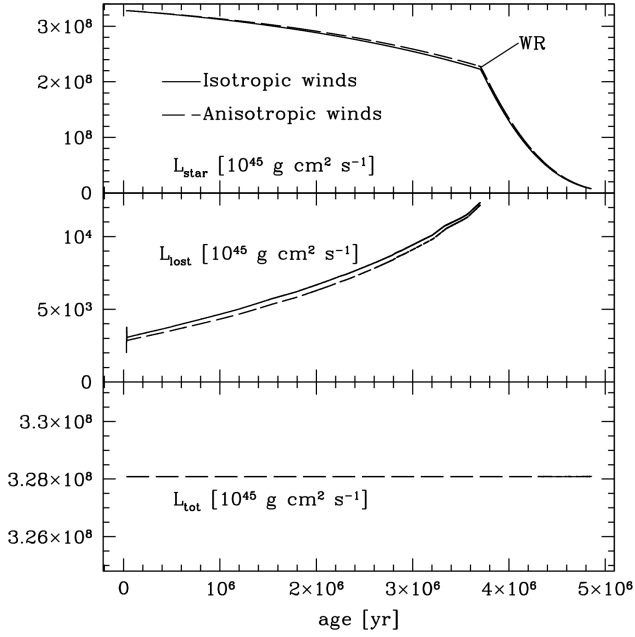}
\caption{Top panel: evolution of the total angular momentum kept in the star (in units of $[10^{45}\,g\,cm^2\,s^{-1}]$). The dased curve is the model with anisotropic wind, the solid curve for the isotropic one. Middle panel: angular momentum removed by wind at each time step, in the same units: dased curve for anisotropic model, solid for isotropic one. Bottom panel: Sum of the angular momentum of the star and the integrated angular momentum removed by wind.}
\label{Levol}
\end{center}
\end{figure}

For both models, the strong mixing induced by the magnetic field leads to the so-called "quasi-chemically-homogenous" evolution (see \citet{Yoon2005a} and \citet{Woosley2006b}). Figure \ref{Levol} (top panel) shows the evolution of the total angular momentum of the stellar interior during its evolution. We see that the anisotropic model (dashed curve) is slightly higher than the isotropic one (solid curve). This is due to the effect of wind anisotropy, as shown in the medium panel: the anisotropic model lose less angular momentum (around $7\%$), while the rotational velocity is high (first part of the evolution). Then, the star becomes a WR, and the high mass loss rate implies a strong breaking of the rotation: the surface velocity is sufficiently below the critical velocity for removing any anisotropy in the wind, and there is no more differences between these two models.

We see that when magnetic field is accounted for, the WR phase has a crucial influence on the total angular momentum kept in the star (and thus in the core, through the coupling produced by the magnetic field). Our model does not keep enough angular momentum for the collapsar model, and is thus not a good GRB progenitor candidate. However, models with very high initial rotational velocity would most probably develop larger differences between the iso-- and anisotropic treatment. The same would be true for models with lower mass loss rate (less massive models, lower metallicity). We will adress this point in a forthcoming paper.

\bibliographystyle{cup}
\bibliography{MyBiblio}

\end{document}